\documentclass[preprint,showpacs,preprintnumbers,amsmath,amssymb,showkeys]{revtex4}

\usepackage{dcolumn}
\usepackage{bm}
\usepackage{amssymb,amsmath}

\begin{document}


\title{Path Integral Solutions to the Distributions of Statistical Mechanics}

\author{Jose A. Magpantay}
\email{jose.magpantay@upd.edu.ph} \affiliation{National Institute
of Physics and Science and Society Program, University of the Philippines, Quezon City 1100,
Philippines\\}
\author{Cilicia Uzziel M. Perez}
\email{uzzie.perez@protonmail.com} \affiliation{Department of Physics and Astronomy, University of Alabama, Tuscaloosa, AL, 35487-0324, USA\\}
\date{\today}

\begin{abstract}
We present the path-integral solutions to the distributions in classical (Gibbs) and quantum (Wigner) statistical mechanics. The kernel of the distributions are derived in two ways - one by time slicing and defining the appropriate short-time interval phase space matrix element and second by making use of the kernel in the path-integral approach to quantum mechanics. We show that the two approaches are perturbatively identical. We also present another computation for the Wigner kernel, which is also the Liouville kernel, for the harmonic oscillator and free particle. These kernels may be used as the starting point in the perturbative expansion of the Wigner kernel for any potential. With the kernel solved, we essentially solve also the distributions in classical and quantum statistical mechanics.    
\end{abstract}

\maketitle

\section{\label{sec:level1}Introduction}
	The distribution function - Gibbs in classical and Wigner in quantum - in statistical mechanics is the starting point in discussing non-equilibrium behavior of physical systems. The distributions are functions of time and phase space variables, coordinates and momenta, and they satisfy partial differential equations that are first-order in time and first order in phase space variables in the classical case and at the minimum quantum correction (order $\hbar^2$), is third order in momenta. There is no general analytic solution for the Wigner function because the equation is at least third-order partial differential equation and most problems are dealt with numerically \cite{Kheruntsyan}. In this paper, we use the path-integral method to present two ways of computing for the kernel of the Wigner equation. To justify the use of the path-integral method, we will summarize the properties of the Wigner distribution and will show what is entailed in solving the Wigner equation.
	
	Quantum mechanics in phase space was formulated by Wigner by defining a 'distribution' in phase space \cite{Wigner} given by
\begin{equation}\label{sha1}
W(x,p;t)=\frac{1}{2\pi\hbar}\int dx' \Psi^*(x+\frac{1}{2}x',t) \Psi(x-\frac{1}{2}x',t) \exp(\frac{i}{\hbar}px').
\end{equation}
Note, the distribution is in quotes because the Wigner distribution is not positive-definite although it satisfies the following
\begin{subequations}\label{sha2}
\begin{gather}
\int W(x,p;t)dp=\left|\Psi(x,t)\right|^2,\label{first}\\
\int W(x,p;t)dx=\left|\Phi(p,t)\right|^2,\label{second}\\
\int W(x,p;t)dxdp=1,
\end{gather}
\end{subequations}
where $\Psi(x,t)$, $\Phi(p,t)$ are the wave functions in coordinate space and momentum space respectively. Wigner showed that W satisfies 
\begin{subequations}\label{sha3}
\begin{gather}
\dfrac{\partial W}{\partial t}+\textbf{H}W=0,\label{first}\\
\textbf{H}=\textbf{L}+2\sum_{n=3}^{odd int} \frac{1}{n!}(\frac{\hbar}{2i})^{n-1}\dfrac{\partial ^n V}{\partial x^n}\dfrac{\partial ^n}{\partial p^n},\label{second}\\
\textbf{L}=\dfrac{\partial V}{\partial x} \dfrac{\partial }{\partial p}-\frac{p}{m} \dfrac{\partial }{\partial x},
\end{gather}
\end{subequations}
The order $\hbar^0$ term of equation (3b) gives the Liouville equation of classical statistical mechanics, which solves for the Gibbs distribution as can be seen from equation(3c). The first quantum correction is of order $\hbar^2$ already, thus it is generally sufficient to only consider the first quantum correction to the Liouville equation in solving for the Wigner distribution. 
	
	The Wigner equation hints of the following form of solution
\begin{equation}\label{sha4}
W(x,p;t)=W_0(x,p;t)+\hbar^2 W_1(x,p;t)+\hbar^4 W_2(x,p;t)+.....
\end{equation}
Any physical quantity $f(x,p)$ will then have a quantum statistical 'expectation value' at any time t given by
\begin{equation}\label{sha5}
\begin{split}
\left\langle f(x,p)\right\rangle_t&=\int dxdp f(x,p) W_0(x,p;t)+\hbar^2\int dxdp f(x,p) W_1(x,p;t)\\
&\quad+\hbar^4\int dxdp f(x,p) W_2(x,p;t)+....
\end{split}
\end{equation}
The first term is the classical statistical expectation value and the quantum corrections are order $\hbar^2$ and higher even orders of $\hbar$. Substituting equation (4) in equation (3), we get the following equations
\begin{subequations}\label{sha6}
\begin{gather}
\left(\dfrac{\partial }{\partial t}+\textbf{L}\right)W_0=0,\label{first}\\
\left(\dfrac{\partial }{\partial t}+\textbf{L}\right)W_1=2\frac{1}{2^2\cdot3!}\dfrac{\partial^3V}{\partial x^3}\dfrac{\partial^3W_0}{\partial p^3},\label{second}\\
\left(\dfrac{\partial }{\partial t}+\textbf{L}\right)W_2=2\frac{1}{2^2\cdot3!}\dfrac{\partial^3V}{\partial x^3}\dfrac{\partial^3W_1}{\partial p^3}-2\frac{1}{2^4\cdot5!}\dfrac{\partial^5V}{\partial x^5}\dfrac{\partial^5W_0}{\partial p^5},
\end{gather}
\end{subequations}
where the equations correspond to even orders in $\hbar$ starting from 0, 2, 4 and so on. Equation (6a) solves for the Gibbs distribution $W_0$ and this is the reason why the first term of equation (5) gives the classical statistical expectation value. The second term, the first quantum correction, is of order $\hbar^2$ already. 

	Given a $W_0$, it means the operator $\left(\dfrac{\partial }{\partial t}+\textbf{L}\right)$ is singular with a square-integrable ($L^2$) zero mode given by
\begin{equation}\label{sha7}
\textit{Z}(x,p;t)=\frac{1}{T^{\frac{1}{2}}}\textbf{W}_0^{\frac{1}{2}},
\end{equation} 
where T, which may eventually be taken to $\infty$, defines the time interval of the system's evolution. Note that the zero mode $\textit{Z}$ has norm 1 because the Gibbs distribution is normalized.	
	
	For equations (6b), (6c) and others to have a solution, we must have the zero mode $\textit{Z}$ orthogonal to the sources $\textsl{S}_i, i = 1, 2,...$ at the right hand side of equations (6b), (6c), etc. This means
\begin{equation}\label{sha8}
\int dxdpdt \textsl{Z}(x,p;t) \textsl{S}_i(x,p;t)=0.
\end{equation}
Furthermore, by equation (2c) and the fact that the Gibbs distribution $W_0$ is normalized, we must have
\begin{equation}\label{sha9}
\int dxdpW_i(x,p;t)=0.
\end{equation}
Equations (8) and (9) are stringent requirements to satisfy, which makes solving the hierarchy of equations given by equations (6;a,b..) difficult to find solutions for. For this reason, we seek another way of solving the Wigner equation and we turn our attention to path-integral methods.
	
	We will solve the Wigner equation via the path-integral method by solving for the kernel of the Wigner equation. As we will show, there are two expressions for this kernel that are not obviously equivalent. One is by direct expansion using the Schrodinger kernel and this we will show in the next section. The other is by direct evaluation of the path-integral by time-slicing and defining an appropriate short time interval matrix element. This we show in Section III. In Section IV, we show that the two perturbative expansions of the kernel are the same term by term. Finally in Section V, we present another solution to the exact kernels for the harmonic oscillator and the free particle. We end by summarizing what was achieved in this work and a discussion of possible applications and open problems in Section VI.    
     	      
\section{\label{sec:level2}The Wigner Kernel in Terms of the Schrodinger Kernel}
	From equation (3a), the Wigner state function is given by
\begin{equation}\label{sha10}
\left|W(t)\right\rangle=\exp{(-\textbf{H}t)}\left|W(0)\right\rangle,
\end{equation}
from which we get the Wigner distribution
\begin{equation}\label{sha11}
W(x,p;t)=\left\langle x,p\left|\exp{(-\textbf{H}t)}\right|W(0)\right\rangle.
\end{equation}
We can define a complete set of states in phase space \cite{Daubechies}, i.e., 
\begin{equation}\label{sha12}
\int dxdp\left|x,p\right\rangle\left\langle x,p\right|=1,
\end{equation}
which gives
\begin{equation}\label{sha13}
W(x,p;t)=\int dx'dp'\textbf{G}(x,p;x',p';t)W(x',p';0),
\end{equation}
where the Wigner kernel is given by
\begin{equation}\label{sha14}
\textbf{G}(x,p;x',p';t)=\left\langle x,p\right|\exp{(-\textbf{H}t)}\left|x',p'\right\rangle.
\end{equation}
Note that we can derive equation (13) directly from equation (1) and in the process derive an expression for the Wigner kernel $\textbf{G}$ in terms of the Schrodinger kernel, see equation (18b) below. This suggests that the completeness relation defined by equation (12) is valid. Furthermore, it must also be true that equation (14) must be equal to equation (18b) and this is one of the results of this paper. 

	We can directly evaluate equation (14) by time-slicing, a procedure used in Feynman's path-integral approach to quantum mechanics, see for example \cite{Feynman}. There, the relevant equations are 
\begin{subequations}\label{sha15}
\begin{gather}
H\Psi=i\hbar \dfrac{\partial \Psi}{\partial t},\label{first}\\
\left|\Psi(t)\right\rangle=\exp{(-\frac{i}{\hbar}Ht)}\left|\Psi(0)\right\rangle,\label{second}\\
\int dx \left|x\right\rangle\left\langle x\right|=1,\label{third}\\
\Psi(x,t)=\int dx' K(x,x';t)\Psi(x',0),\label{fourth}\\
K(x,x';t)=\left\langle x\right|\exp{(-\frac{i}{\hbar}Ht)}\left|x'\right\rangle.
\end{gather}
\end{subequations}
Time slicing yields the path-integral
\begin{equation}\label{sha16}
K(x,x';t)=\int_{x(0)=x',x(t)=x} [dx(t)] \exp{(\frac{i}{\hbar}\int L(x,\dot{x})dt)},
\end{equation}
where $L(x,\dot{x})$ is the Lagrangian of the system.
For the Wigner kernel, there is no closed form expression that corresponds to equation (16), we can only present an expression that should give the analogue of the Lagrangian in equation (16) provided we can carry out certain path integrations, see equation (30) of the next section. However, we will derive a perturbation expansion and show that it is the same as the perturbation expansion derived using the method given below. 

	As already stated above, there is another path-integral solution which will directly relate $\textbf{G}(x,p;x'p';t)$ to $K(x,x';t)$. Using equation (15d) in equation (1), we find
\begin{equation}\label{sha17}
W(x,p;t)=\frac{1}{2\pi\hbar}\int dx'\exp^{(\frac{i}{\hbar}px')}\int dydy'K^*(x+\frac{x'}{2},y;t)K(x-\frac{x'}{2},y';t)\rho(y,y';0),
\end{equation}
where $\rho(y,y';0)$ is the density matrix at $t=0$. We can solve for $\rho(y,y';0)$ in terms of $W(x,p;0)$ to give
\begin{subequations}\label{sha18}
\begin{gather}
W(x,p;t)=\int dx'dp'\textbf{G}(x,p;x',p';t)W(x',p';0),\label{first}\\
\textbf{G}=\frac{1}{2\pi \hbar}\int dsds'\exp{(\frac{i}{\hbar}ps)}K(x-\frac{s}{2},x'-\frac{s'}{2};t)K^*(x+\frac{s}{2},x'+\frac{s'}{2};t)\exp{(-\frac{i}{\hbar}p's')}.
\end{gather}
\end{subequations}
Equation (18b) was discussed in the time-independent case in \cite{Moshinsky} and later extended to the time-dependent in \cite{Schuch} for the free particle and harmonic oscillator cases and then explicitly derived in general in \cite{Sels}. Equations (14) and (18b) are the two expressions for the Wigner kernel. 

	A perturbative expansion can be given for equation (18b) by making use of the perturbation expansion for the Schrodinger kernel K. To clarify how the expansion is to be made, let us write the Lagrangian first as
\begin{subequations}\label{sha19}
\begin{gather}
L=L_0-\tilde{V},\label{first}\\
L_0= \frac{1}{2}mv^2-V_0(x),
\end{gather}
\end{subequations}
where $V_0(x)$ is a potential with a known, exact kernel $K_0$. Examples of these are
\begin{subequations}\label{sha20}
\begin{gather}
K_0^{fp}(x,x';t)=(\frac{m}{2\pi i\hbar t})^{\frac{1}{2}}\exp{\left[\frac{im}{2\hbar t}(x-x')^2\right]},\label{first}\\
K_0^{ho}(x,x';t)=(\dfrac{\omega}{2\pi i \hbar \sin \omega t})^{\frac{1}{2}}\exp{\left\{\frac{i\omega}{2\hbar \sin \omega t}\left[(x^2+x'^2)\cos \omega t-2xx'\right]\right\}},
\end{gather}
\end{subequations}
where $K_0^{fp}$ is the Schrodinger kernel for the free particle while $K_0^{ho}$ is the kernel for the harmonic oscillator with potential $V_0=\frac{1}{2}\omega^2x^2$ and unit mass.
Substituting equation (19) in the path integral given by equation (16), we get the perturbation expansion for the kernel K given by
\begin{equation}\label{sha21}
\begin{split}
K(x,x';t)&=K_0(x,x';t)+(\frac{-i}{\hbar})\int dx_adt_aK_0(x,x_a;t-t_a)\tilde{V}(x_a)K_0(x_a,x';t_a)\\
&\quad+(\frac{-i}{\hbar})^2\int \int dx_adt_adx_bdt_bK_0(x,x_a;t-t_a)\tilde{V}(x_a)K_0(x_a,x_b;t_a-t_b)\\
&\quad \times \tilde{V}(x_b)K_0(x_b,x';t_b)+....
\end{split}
\end{equation}
Substituting equation (21) in equation (18b) gives the perturbative expansion for the Wigner kernel. The perturbation expansion for $\textbf{G}$ begins with an exact and closed form $\textbf{G}_0$, which is solved from $K_0$ using equation (18b). For the free particle and harmonic oscillator, these are
\begin{subequations}\label{sha22}
\begin{gather}
\textbf{G}_0^{fp}=\frac{m}{t}\delta(p-\frac{m}{t}(x-x'))\delta(p'-\frac{m}{t}(x-x')),\label{first}\\
\textbf{G}_0^{ho}=\frac{\omega}{\sin(\omega t)}\delta\left(p-\omega x\cot(\omega t)+\frac{\omega x'}{\sin(\omega t)}\right)\delta\left(p'+\omega x'\cot(\omega t)-\frac{\omega x}{\sin(\omega t)}\right).
\end{gather}
\end{subequations}
As a check, we see that equation (22a) with unit mass follows from equation (22b) as $\omega \rightarrow 0$. Furthermore, we can cross check equation (22b) with the results of \cite{Kim} for the harmonic oscillator, say in the ground state, where the Wigner distribution was shown to be a constant proportional to the exponential of the ground state energy. Equation (22b) yields the same result.

	Naively, the higher order terms in equation (21) seems to give negative powers in $\hbar$ for the $\textbf{G}$ expansion. This is not what we expect from the discussions in the previous section. We will resolve this later, i.e., we will show that the perturbative expansion of $\textbf{G}$ based on equations (18b) and equation (21) is really an expansion in positive powers of $\hbar$. And more importantly, we will show that this expansion is the same as the perturbation expansion that will be derived in another way in the next section. 
\section{\label{sec:level3}Another Perturbation Expansion}
	Equation (14) gives an expression for the Wigner kernel reminiscent of the Schrodinger kernel given by equations (15e). But here we face the important question of what the state $\left|x,p\right\rangle$ means because in phase space we do not have simultaneous eigenstates of position and momentum. We will not settle this issue but will propose a matrix element $\left\langle x_i,p_i\right|x_{i-1},p_{i-1}>$, which is consistent with the uncertainty principle and at the same time allow the evaluation of the Wigner kernel that will give the same result as equation (18b). Just like in deriving equation (16) from equation (15e) in the Schrodinger-Feynman quantum mechanics, we will also do time-slicing, i.e., divide the time interval [0,t] into N infinitesimal intervals of $\epsilon=t/N$, i.e., $t-t_N=t_N-t_{N-1}=....=t_1-0$. Then, we insert a complete set of states as given by equation (12), which apparently is justified even in phase space. The Wigner kernel then becomes
\begin{equation}\label{sha23}
\begin{split}
\textbf{G}(x,p;x',p';t)&=\int dx_Ndp_N....dx_1dp_1 \left\langle x,p\right|\exp{(-\textbf{H}\epsilon)}\left|x_N,p_N\right\rangle...\\
&\quad \cdot\left\langle x_i,p_i\right|\exp{(-\textbf{H}\epsilon)}\left|x_{i-1},p_{i-1}\right\rangle...\left\langle x_1,p_1\right|\exp{(-\textbf{H}\epsilon)}\left|x',p'\right\rangle.
\end{split}
\end{equation}
The Wigner kernel requires the evaluation of the short time interval matrix element STIME given by
\begin{equation}\label{sha24}
\begin{split}
STIME&=\left\langle x_i,p_i\right|{1-\epsilon[(\dfrac{\partial V_0}{\partial x} \dfrac{\partial }{\partial p}-\frac{p}{m} \dfrac{\partial }{\partial x})+\dfrac{\partial \tilde{V}}{\partial x}\dfrac{\partial }{\partial p}-\frac{\hbar^2}{12}\dfrac{\partial^3 \tilde{V}(x)}{\partial x^3} \dfrac{\partial^3}{\partial p^3}]}+..\left|x_{i-1},p_{i-1}\right\rangle\\
&=F(x_i,p_i;x_{i-1},p_{i-1})-\epsilon[(\dfrac{\partial V_0}{\partial x_i} \dfrac{\partial }{\partial p_{i-1}}-\frac{p_i}{m} \dfrac{\partial }{\partial x_{i-1}})F(x_i,p_i;x_{i-1},p_{i-1})\\
&\quad+(\dfrac{\partial \tilde{V}}{\partial x_i}\dfrac{\partial }{\partial p_{i-1}}-\frac{\hbar^2}{12}\dfrac{\partial^3 \tilde{V}}{\partial x_i^3} \dfrac{\partial^3}{\partial p_{i-1}^3})F(x_i,p_i;x_{i-1},p_{i-1})]+...,
\end{split}
\end{equation}
where 
\begin{equation}\label{sha25}
F(x_i,p_i;x_{i-1},p_{i-1})=\left\langle x_i,p_i|x_{i-1},p_{i-1}\right\rangle .
\end{equation}
Substituting equation (24) in equation (23), it is clear that the analog of equation (16) in the path-integral formulation of quantum mechanics is not straightforward to derive. To compare with quantum mechanics, the analogue expression of $F(x,p;x',p')$ is simply $F(x,x')=\left\langle x|x'\right\rangle = \delta(x-x')$. In phase space, $F(x,p;x',p')$ is restricted by the uncertainty principle, which limits the specification of both position and momentum at the same time. In quantum mechanics, the short-time interval matrix element does not involve derivatives of the coordinates while the STIME in the Wigner kernel involve derivatives in phase space. This will not allow the crucial step used in quantum mechanics, the exponentiation of the short time intervals yielding in equation (16). For these reasons, the evaluation of the Wigner kernel is more involved.

	It is clear from equation (24) that to be able to exponentiate the STIME factors in equation (23), we must remove the differentiations in phase space variables. We will do this guided by the uncertainty principle. We will take 
\begin{equation}\label{sha26}
F(x_i,p_i;x_{i-1},p_{i-1})=f(x_i-x_{i-1})g(p_i-p_{i-1}),
\end{equation}
where f and g are Gaussian functions given by
\begin{subequations}\label{sha27}
\begin{gather}
f(x_i-x_{i-1})=\frac{1}{\sqrt{\pi \epsilon a}}\exp{[-\frac{1}{\epsilon a}(x_i-x_{i-1})^2]},\label{first}\\
g(p_i-p_{i-1})=\frac{1}{\sqrt{\pi \epsilon b}}\exp{[-\frac{1}{\epsilon b}(p_i-p_{i-1})^2]},
\end{gather}
\end{subequations}		
where a, b are arbitrary (dimensional) parameters. The use of Gaussian functions clearly will satisfy the uncertainty principle and as the following discussions and Section V will show, we will be able to evaluate perturbatively the Wigner kernel. The inclusion of $\epsilon$, the small-time interval in equation (23), will be clear in the next few lines. The uncertainty principle restricts a and b to satisfy $ab \epsilon^2=\hbar^2$.
We then write these expressions as integrals given by
\begin{subequations}\label{sha28}
\begin{gather}
f(x_i-x_{i-1})=\frac{1}{\pi}\int_{-\infty}^\infty d\alpha_i \exp[i\alpha_i(x_i-x_{i-1})-\frac{\epsilon a}{4}\alpha_i^2],\label{first}\\
g(p_i-p_{i-1})=\frac{1}{\pi}\int_{-\infty}^\infty d\beta_i \exp[i\beta_i(p_i-p_{i-1})-\frac{\epsilon b}{4}\beta_i^2].
\end{gather}
\end{subequations}
Using equation (28) in equation (24), we find
\begin{equation}\label{sha29}
\begin{split}
STIME&=\int d\alpha_i d\beta_i \left\{1-\epsilon\left[\dfrac{\partial V}{\partial x_i}(i\beta_i)-\frac{p_i}{m}(-i\alpha_i)+\frac{\hbar^2}{12}\dfrac{\partial^3V}{\partial x_i^3}(-i\beta_i)^3\right]+...\right\}\\
&\quad\times\exp[i\alpha_i(x_i-x_{i-1})+i\beta_i(p_i-p_{i-1})-\frac{\epsilon a}{4}\alpha_i^2-\frac{\epsilon b}{4}\beta_i^2].
\end{split}
\end{equation}
Substituting equation (29) in equation (23), we find
\begin{equation}\label{sha30}
\begin{split}
\textbf{G}(x,p;x',p';t)&=\int(dx)(dp)(d\alpha)(d\beta)\exp \left\{\int dt\left[-\frac{a}{4}\alpha^2(t)-\frac{b}{4}\beta^2(t)\right.\right.\\
&\quad +i\alpha(t)(\frac{p(t)}{m}+\dot{x}(t))+i\beta(t)(\dfrac{\partial V_0}{\partial x}+\dot{p}(t))\\
&\left.\left.\quad+i\beta(t)\dfrac{\partial \tilde{V}}{\partial x}+i\frac{\hbar^2}{12}\beta^3(t)\dfrac{\partial^3\tilde{V}}{\partial x^3}\right]\right\}.
\end{split}
\end{equation}
	If we can integrate out the $\alpha$ and $\beta$ path integrals above, we will get the analogous expression to equation (16) of the Schrodinger kernel for the Wigner kernel $\textbf{G}$ and derive the counterpart of the Lagrangian (in this case function of both (x,p)) for the Wigner equation. The $\alpha$ term is quadratic and is trivially integrable. The $\beta$ term poses two problems - first it is cubic and thus not exactly integrable. Second, even with just the linear term $\beta(t)(\dfrac{\partial V}{\partial x}+\dot{p}(t))$, upon integration yields a term $\propto (\dfrac{\partial V}{\partial x}+\dot{p}(t))^2$ and if V is of order $x^3$, the path integral in x is not exact and the phase space path integral will not be exact. This is where the decomposition given by equations (19a) and (19b), which is already accounted for in equation (30), becomes important for this will enable us to isolate the exact $\textbf{G}_0$ and expand $\textbf{G}$ perturbatively.
\begin{equation}\label{sha31}
\begin{split}
\textbf{G}_0&=\int(dx)(dp)(d\alpha)(d\beta)\exp\left\{\int dt\left[-\frac{a}{4}\alpha^2(t)-\frac{b}{4}\beta^2(t)\right.\right.\\  
&\left.\left.\quad +i\alpha(t)(\frac{p(t)}{m}+\dot{x}(t))+i\beta(t)(\dfrac{\partial V_0}{\partial x}+\dot{p}(t))\right]\right\}.
\end{split}
\end{equation}
This is exactly integrable for $V_0$ given by the free particle and harmonic oscillator. As we will show in Section V, this expression gives the Wigner kernel, which in these cases are the same as the Liouville kernel, given by equations (22a) and (22b). 
	
	Now we can expand $\textbf{G}(x,p;x',p';t)$ by expanding the exponential term $\exp\left\{\int dt[i\beta(t)\dfrac{\partial \tilde{V}}{\partial x}+i\frac{\hbar^2}{12}\beta^3(t)\dfrac{\partial^3V}{\partial x^3}]\right\}$ explicitly up to two terms and noting that $i\beta(t)$ arose from $\dfrac{\partial}{\partial p(t)}$, we find	    
\begin{equation}\label{sha32}
\begin{split}
\textbf{G}(x,p;x',p';t)&=\textbf{G}_0(x,p;x',p';t)+\int dx_adp_adt_a\textbf{G}_0(x,p;x_a,p_a;t-t_a)\dfrac{\partial \tilde{V}(x_a)}{\partial x_a}\dfrac{\partial}{\partial p_a}\textbf{G}_0(x_a,p_a;x',p';t_a)\\
&\quad +\frac{\hbar^2}{12}\int dx_adp_adt_a\textbf{G}_0(x,p;x_a,p_a;t-t_a)\dfrac{\partial^3\tilde{V}}{\partial x_a^3}\dfrac{\partial^3}{\partial p_a^3}\textbf{G}_0(x_a,p_a;x',p';t_a)+...
\end{split}
\end{equation}
This equation is the perturbation expansion of the Wigner kernel, which is defined by equation (14). The first and second terms give the Liouville kernel. The third term gives the quantum correction to yield the Wigner kernel. 

	Let us spell out the corresponding equations in coordinate space quantum mechanics and phase space quantum mechanics. In coordinate space quantum mechanics, the equations are defined by equations (15a) to (15e) and equation (16) for the path-integral expression for the kernel. In phase space quantum mechanics, the corresponding equations are equations (3), (10), (12), (18a), (14) respectively and the corresponding path-integral expression for the Wigner kernel is equation (30). The perturbation expansion for the Wigner kernel given by equation (32) corresponds to the perturbation expansion for the Schrodinger kernel given by equation (21). 
	
	But the two kernels are related by equation (18b) and it is not obvious at all that using equation (21) in equation (18b) will yield the same result as equation (32). This is what we will show in the next section.   	           	 
\section{\label{sec:level4}Equivalence of the Two Perturbation Expansions}
	As already pointed out in Section II, equations (21) and (18b) seem to give an expansion with the wrong powers in $\hbar$. Now we will show how the right powers in $\hbar$ arises and that the expansion is precisely given by equation (32). 
	The first term of equation (21) substituted in equation (18b) gives $\textbf{G}_0$. The next in the expansion, which is of order $\frac{i}{\hbar}$, has two terms and it is these terms that give the second and third terms of equation (34).
\begin{equation}\label{sha33}
\begin{split}
\textbf{N}&=\frac{i}{\hbar}\int dsds'\int dx_adt_a\exp(\frac{i}{\hbar}ps)\Big[  K_0(x-\frac{s}{2},x'-\frac{s'}{2};t)K_0^*(x+\frac{s}{2},x_a;t-t_a)\tilde{V}(x_a)K_0^*(x_a,x'+\frac{s'}{2};t_a) \\
&\quad -K_0(x-\frac{s}{2},x_a;t-t_a)\tilde{V}(x_a)K_0(x_a,x'-\frac{s'}{2};t_a)K_0^*(x+\frac{s}{2},x'+\frac{s'}{2};t)\exp(-\frac{i}{\hbar}p's')\Big]
\end{split}
\end{equation}
This term only has 3 $K_0's$ while the second and third terms of equation (32), because they have 2 $\textbf{G}_0's$, has 4 $K_0's$. This suggests the use of the following property of the Schrodinger kernel 
\begin{equation}\label{sha34}
K(x,x';t)=\int dx_b K(x,x_b;t-t_b)K(x_b,x';t_b).
\end{equation}
Using this in appropriate parts in equation (33) and redefining integration variables 
\begin{equation}\label{sha35}
\begin{split}
\textbf{N}&=\frac{i}{\hbar}\int dsds'dx_adt_adr'\exp{(\frac{i}{\hbar}ps)}K_0(x-\frac{s}{2},x_a-\frac{s'}{2};t-t_a)K_0^*(x+\frac{s}{2},x'+\frac{s'}{2};t-t_a)\\
&\quad \times \left[\tilde{V}(x_a-\frac{s'}{2})-\tilde{V}(x_a+\frac{s'}{2})\right]K_0(x_a-\frac{s'}{2},x'-\frac{r'}{2};t_a)K_0^*(x_a+\frac{s'}{2},x'+\frac{r'}{2};t_a)\exp{(-\frac{i}{\hbar}p'r')}
\end{split}
\end{equation}
Expanding $\tilde{V}$, we get only odd powers in s' with odd powers of derivatives of $\tilde{V}(x_a)$. This suggests the expansion given in equation (32), only we have to figure out how the odd powers in momentum derivatives acting on $\textbf{G}_0$ appear. This happens by writing down, for example for the term linear in s', 
\begin{equation}\label{sha36}
\begin{split}
s'K_0(x_a-\frac{s'}{2},x'-\frac{r'}{2};t_a)K_0^*(x_a+\frac{s'}{2},x'+\frac{r'}{2};t_a)&=\frac{\hbar}{i}\int dp_a\exp{(-\frac{i}{\hbar}p_as')}\dfrac{\partial}{\partial p_a}\int dr\exp{(\frac{i}{\hbar}p_ar)}\\
&\quad \times K_0(x_a-\frac{r}{2},x'-\frac{r'}{2};t_a)K_0^*(x_a+\frac{r}{2},x'+\frac{r'}{2};t_a).
\end{split}
\end{equation}
Expressed this way, we clearly see that the linear term in s' is precisely the second term of equation (32). We also see how the positive powers of $\hbar$ appear. The $s'^3$, which goes with $\dfrac{\partial^3\tilde{V}}{\partial x_a^3}$ gives the third term of equation (32) with the right factor of $\frac{\hbar^2}{12}$ by following the same decomposition.

	What we have explicitly shown is the equivalence of equation (32) with the first and second terms of the expansion of the Wigner kernel using equation (18b) and the expansion of the Schrodinger kernel given by equation (21). But what about the terms represented by ellipsis in equation (32)? There are terms of order $\hbar^n$ with $n=4,6,...$ coming from including the other terms in the Wigner equation as found in equation (3b). Following the steps in equations (23), (27), (29) and (31) will result in terms such as 
\begin{equation}\label{sha37}
\hbar^n\int dx_adp_adt_a\textbf{G}_0(x,p;x_a,p_a;t_a)\dfrac{\partial^{n+1}\tilde{V}}{\partial x_a^{n+1}}\dfrac{\partial^{n+1}}{\partial p_a^{n+1}}\textbf{G}_0(x_a,p_a;x',p';t-t_a),
\end{equation}
with $n=4,6,...$. These terms still follow from equations (37) by expanding $[\tilde{V}(x_a-\frac{s'}{2})-\tilde{V}(x_a+\frac{s'}{2})]$ up to $s'^m$ with $m=5,7,...$ (note the even powers of m cancel) by following equation (36). 

	But all these terms (the explicit three terms of (32) and (37)) arise from the $\frac{i}{\hbar}$ term of equation (21) and equation (18b). The next term in the expansion of (18b) using equation (21) is order $(\frac{1}{\hbar})^2$ containing 4 factors of $K_0$ and two factors of $\tilde{V}$. This term, as we will show below,  is the same as the term from equation (32) by expanding $\exp\left\{\int dt[i\beta(t)\dfrac{\partial \tilde{V}}{\partial x}+i\frac{\hbar^2}{12}\beta^3(t)\dfrac{\partial^3V}{\partial x^3}]\right\}$ up to the third term. This part of the expansion of equation (30) has four terms - terms of order $\hbar^0$, $\hbar^2$ and $\hbar^4$. The $O(\hbar^0)$ term is
\begin{equation}\label{sha38}
\int dx_adp_adt_adx_bdp_bdt_b\textbf{G}_0(x,p;x_a,p_a;t-t_a)\dfrac{\partial \tilde{V}}{\partial x_a}\dfrac{\partial}{\partial p_a}\textbf{G}_0(x_a,p_a;x_b,p_b;t_a-t_b)\dfrac{\partial \tilde{V}}{\partial x_b}\dfrac{\partial}{\partial p_b}\textbf{G}_0(x_b,p_b;x',p';t_b).
\end{equation}
This is clearly the leading correction to the Liouville kernel given by the first and second terms of equation (32). The $O(\hbar^2)$ terms are given by
\begin{equation}\label{sha39}
\int dx_adp_adt_adx_bdp_bdt_b\textbf{G}_0(x,p;x_a,p_a;t-t_a)\dfrac{\partial \tilde{V}}{\partial x_a}\dfrac{\partial}{\partial p_a}\textbf{G}_0(x_a,p_a;x_b,p_b;t_a-t_b)\dfrac{\partial^3 \tilde{V}}{\partial x_b^3}\dfrac{\partial^3}{\partial p_b^3}\textbf{G}_0(x_b,p_b;x',p';t_b),
\end{equation}
and a similar term with the third derivatives ahead of the first derivatives. This is still $O(\hbar^2)$ and is a higher order correction to the Wigner correction to the Liouville kernel given by the third term of equation (32). Finally, the $O(\hbar^4)$ term is
\begin{equation}\label{sha40}
\int dx_adp_adt_adx_bdp_bdt_b\textbf{G}_0(x,p;x_a,p_a;t-t_a)\dfrac{\partial^3 \tilde{V}}{\partial x_a^3}\dfrac{\partial^3}{\partial p_a^3}\textbf{G}_0(x_a,p_a;x_b,p_b;t_a-t_b)\dfrac{\partial^3 \tilde{V}}{\partial x_b^3}\dfrac{\partial^3}{\partial p_b^3}\textbf{G}_0(x_b,p_b;x',p';t_b).        \end{equation}
This is a higher order correction to the contribution given by equation (37) for $n=4$.

	Now we will show that the corrections given by equations (38), (39) and (40) are the same as the $O((\frac{1}{\hbar})^2)$ term of the expansion of equation (18b) using equation (21).  There are three such terms with 4 space integrals (over $(s,s',x_a,x_b)$) and two time integrals (over $(t_a,t_b)$) involving either $K_0K_0^*\tilde{V}K_0^*\tilde{V}K_0^*, K_0\tilde{V}K_0\tilde{V}K_0K_0^*,K_0\tilde{V}K_0K_0^*\tilde{V}K_0^*)$. Using equation (34) twice appropriately in each term and redefining integration variables, we find the terms simplify to a term with  $K_0K_0^*[\tilde{V}(x_a+\frac{s'}{2})-\tilde{V}(x_a-\frac{s'}{2})]K_0K_0^*[\tilde{V}(x_b+\frac{r'}{2})-\tilde{V}(x_b-\frac{r'}{2})]K_0K_0^*$. Expanding $\tilde{V}$, we find only terms with odd derivatives. Then we use equation (36) twice appropriately, we are able to introduce two momenta integrals and we clearly see that this term precisely gives equations (38) to (40). 

	The other terms in the expansion of equation (18b) and (21) also give the expansion of equation (30). This proves the equivalence of the two perturbation expansions. 
	
	If we gather all the terms in equations (32), (37) to (40), we have an expression for the Wigner kernel containing quantum corrections up to $\hbar^4$. Further higher order corrections will appear when we expand $\exp\left\{\int dt[i\beta(t)\dfrac{\partial \tilde{V}}{\partial x}+i\frac{\hbar^2}{12}\beta^3(t)\dfrac{\partial^3V}{\partial x^3}]\right\}$ to include fourth and higher terms in the expansion. Putting together all the $\hbar^0$ terms will give the Liouville kernel given by
\begin{equation}\label{sha41}
\begin{split}
\textbf{G}_L(x,p;x',p';t)&=\textbf{G}_0(x,p;x',p';t)+\int dx_adp_adt_a\textbf{G}_0(x,p;x_a,p_a;t-t_a)\dfrac{\partial \tilde{V}(x_a)}{\partial x_a}\dfrac{\partial}{\partial p_a}\textbf{G}_0(x_a,p_a;x',p';t_a)\\
&\quad -\frac{1}{2}\int dx_adp_adt_adx_bdp_bdt_b\textbf{G}_0(x,p;x_a,p_a;t-t_a)\dfrac{\partial \tilde{V}}{\partial x_a}\dfrac{\partial}{\partial p_a}\\
&\quad \textbf{G}_0(x_a,p_a;x_b,p_b;t_a-t_b)\dfrac{\partial \tilde{V}}{\partial x_b}\dfrac{\partial}{\partial p_b}\textbf{G}_0(x_b,p_b;x',p';t_b)+....
\end{split}
\end{equation}
Including all the $\hbar^2$ terms will give the Wigner kernel up to this order given by
\begin{equation}\label{sha42}
\textbf{G}(x,p;x',p';t)=\textbf{G}_L(x,p;x',p';t)+\frac{\hbar^2}{12}\int dx_adp_adt_a\textbf{G}_L(x,p;x_a,p_a;t-t_a)\dfrac{\partial^3\tilde{V}}{\partial x_a^3}\dfrac{\partial^3}{\partial p_a^3}\textbf{G}_L(x_a,p_a;x',p';t_a)
\end{equation}
Using equation (41) or (42) in equation (18a), we solve for the Liouville or Wigner distribution given the initial distribution. This means we have solved the distributions of statistical mechanics, classical and quantum mechanical.
\section{\label{sec:level5}Another Derivation of the Exact Wigner Kernels}
	We will show how to evaluate equation (31) in the cases of the free particle and harmonic oscillator to give another derivation of equations (22a) and (22b), which can be derived using equation (18b) and equations (20a,b). We will spell out the details of the computation to show that the methods defined in Section III, in particular equations (26) and (27) are valid by rederiving a known result. 
Integrating $\alpha(t)$ and $\beta(t)$, we get
\begin{equation}\label{sha43}
\textbf{G}_0=\left(\dfrac{1}{ab\epsilon^2}\right)^{(\frac{N+1}{2})}\int (dx)(dp)\exp{\left\{\int dt [-\frac{1}{a}(\frac{p}{m}-\dot{x})^2-\frac{1}{b}(\dot{p}+\dfrac{\partial V_0}{\partial x})^2]\right\}}.
\end{equation}
We will evaluate this path-integral by treating the integrand as an effective `Lagrangian' in phase space, i.e.,
\begin{equation}\label{sha44}
L_{eff}(x,p)=-\frac{1}{a}(\frac{p}{m}-\dot{x})^2-\frac{1}{b}(\dot{p}+\dfrac{\partial V_0}{\partial x})^2.
\end{equation}
We will use the stationary phase approximation, which in this case will give exact results, for the harmonic oscillator and free particle. Consider first the harmonic oscillator, having unit mass and potential $V_0=\frac{1}{2}\omega x^2$. We write
\begin{subequations}\label{sha45}
\begin{align}
x=x_{cl}+\delta x,\label{first}\\
p=p_{cl}+\delta p,
\end{align}
\end{subequations}
where $x_{cl}$ and $p_{cl}$ refer to the classical solutions of $L_{eff}$. The kernel becomes
\begin{equation}\label{sha46}
\textbf{G}_0=\left(\dfrac{1}{ab\epsilon^2}\right)^{(\frac{N+1}{2})}\exp[{\int_0^t dt' L_{eff}(x_{cl}(t'),p_{cl}(t'))}] \int (d\delta x)(d\delta p)\exp({-\int dt \delta^T M \delta}),
\end{equation}
where the matrix elements are given by
\begin{subequations}\label{sha47}
\begin{align}
\delta=\begin{pmatrix}
\delta x\\ 
\delta p\\
\end{pmatrix},\label{first}\\
M=\begin{pmatrix}
\kappa & \phi \\
\gamma & \rho \\
\end{pmatrix},
\end{align}
\end{subequations}
with the elements of the M matrix given by
\begin{subequations}\label{sha48}
\begin{gather}
\kappa=-\frac{1}{a}\dfrac{d^2}{dt^2}+\frac{1}{b}\omega^4,\label{first}\\
\phi=\frac{2}{b}\omega^2\dfrac{d}{dt},\label{second}\\
\gamma=-\frac{2}{a}\dfrac{d}{dt},\label{third}\\
\rho=-\frac{1}{b}\dfrac{d^2}{dt^2}+\frac{1}{a}.
\end{gather}
\end{subequations} 
 
 To get the first factor of equation (46), we need to get the relevant classical solutions. The Euler-Lagrange equation of equation (44) are given by
\begin{subequations}\label{sha49}
\begin{gather}
\dfrac{du}{dt}+\omega^2v=0,\label{first}\\
\dfrac{dv}{dt}-u=0,\label{second}\\
u=\frac{1}{a}(p-\dot{x}),\label{third}\\
v=\frac{1}{b}(\dot{p}+\omega^2x).
\end{gather}
\end{subequations}
It is clear that the solution of Hamilton's equations, $p-\dot{x}=0$, $\dot{p}+\omega^2 x=0$ are solutions of the above equations. These equations give
\begin{equation}\label{sha50}
x_H(t')=A\cos(\omega t')+B\sin(\omega t'),
\end{equation}
and using the end points conditions $x(0)=x'$, $x(t)=x$, we find
\begin{subequations}\label{sha51}
\begin{gather}
A=x',\label{first}\\
B=x\dfrac{1}{\sin(\omega t)}-x'\cot(\omega t).
\end{gather}
\end{subequations}
But the Hamilton equations also give $p_H(t')=\dot{x}$ and using the end points conditions give
\begin{subequations}\label{sha52}
\begin{gather}
p'=x\dfrac{\omega}{\sin(\omega t)}-x'\omega \cot(\omega t),\label{first}\\
p=-x'\dfrac{\omega}{\sin(\omega t)}+x\omega \cot(\omega t),
\end{gather}
\end{subequations}
from which we get the constraints
\begin{subequations}\label{sha53}
\begin{gather}
\chi_1=p+x'\dfrac{\omega}{\sin(\omega t)}-x\omega \cot(\omega t) \approx 0,\label{first}\\
\chi_2=p'-x\dfrac{\omega}{\sin(\omega t)}+x'\omega \cot(\omega t) \approx 0,
\end{gather}
\end{subequations}
where $\approx$ is taken as weakly equal to zero a la Dirac. 

	Other than the Hamilton equation solutions, is there another solution to equations (49)? The answer is yes and it is this solution that will give the first factor of equation (46). Differentiating equation (49) gives u and v both satisfying harmonic equations with solutions
\begin{subequations}\label{sha54}
\begin{gather}
u(t')=C\cos(\omega t')+D\sin(\omega t'),\label{first}\\
v(t')=E\cos(\omega t')+F\sin(\omega t').
\end{gather}
\end{subequations}
Using the end point conditions, we can express the integration constants in terms of the constraints $\chi_1$ and $\chi_2$ as
\begin{subequations}\label{sha55}
\begin{gather}
C=\chi_2,\label{first}\\
D=\dfrac{1}{\sin\omega t}\chi_1,\label{second}\\
E=\dfrac{\omega}{b\sin\omega t}\chi_1,\label{third}\\
F=-\omega \chi_2.
\end{gather}
\end{subequations}
Substituting equations (54) and (55) in (44) and integrating out t' in the first factor of (46), we find  
\begin{equation}\label{sha56}
\int_0^t dt'L_{eff}(x_{cl}(t'),p_{cl}(t'))=-\frac{1}{a}\textit{F}_1^2-\frac{1}{b}\textit{F}_2^2,
\end{equation}
where $\textit{F}_1^2$ and $\textit{F}_2^2$ are quadratic functions of $\chi_1$ and $\chi_2$ and polynomials of t, the detailed forms of each are not important.

	The path-integral of the fluctuations give $det^{-\frac{1}{2}}M$, where M is given by equations (47) and (48). Let us write the matrix M as
\begin{equation}\label{sha57}
M=M_0+M_c,
\end{equation}
where
\begin{subequations}\label{sha58}
\begin{gather}
M_0=\begin{pmatrix}
-\frac{1}{a}(\dfrac{d^2}{dt^2}+\omega^2) & 0 \\
0 & -\frac{1}{b}(\dfrac{d^2}{dt^2}+\omega^2) \\
\end{pmatrix},\label{first}\\
M_c=\begin{pmatrix}
\frac{\omega^2}{a}+\frac{\omega^4}{b} & \phi \\
\gamma & \frac{1}{a}+\frac{\omega^2}{b} \\
\end{pmatrix}.
\end{gather}
\end{subequations}
The determinant is given by
\begin{subequations}\label{sha59}
\begin{gather}
detM= detM_0 det\left(\textbf{1}+M_0^{-1}M_c\right),\label{first}\\
detM_0=det\left(-\frac{1}{a}(\dfrac{d^2}{dt^2}+\omega^2)\right)det\left(-\frac{1}{b}(\dfrac{d^2}{dt^2}+\omega^2)\right).
\end{gather}
\end{subequations}
We now show that
\begin{equation}\label{sha60}
det^{-\frac{1}{2}}\left(-\frac{1}{a}(\dfrac{d^2}{dt^2}+\omega^2)\right)=a^{\frac{N}{2}}\left(\dfrac{\omega}{2\pi \sin(\omega t)}\right)^{\frac{1}{2}}.
\end{equation}
The factor $a^{\frac{N}{2}}$ easily follows by writing the path-integral expression, do the time slicing and noting that $(d\delta x(t'))=d\delta x(t_1)....d\delta x(t_N)$ because the end points $\delta x(t'=0)$ and $\delta x(t'=t_{N+1}=t)$ are fixed with zero values. By scaling the fluctuations from $\delta x$ to $\frac{1}{\sqrt{a}}\delta x$, we get the factor $a^{\frac{N}{2}}$. The other factor in equation(50) is standard harmonic oscillator term. 

	Taking into account equation (56) and (60) in (46), we see that $\textbf{G}_0$ has the following leading terms (we will gather the remaining terms at the end and argue that they can be absorbed in the normalization)
\begin{equation}\label{sha61}
\textbf{G}_0=\left(\dfrac{\omega}{2\pi \sin(\omega t)}\right)\frac{1}{\sqrt{a}\sqrt{b}}\exp{(-\frac{1}{a}\textit{F}_1^2-\frac{1}{b}\textit{F}_2^2)}.
\end{equation}
The uncertainty principle discussion in Section III, right before equation (28), says that $ab \epsilon^2=\hbar^2$. Let us estimate the dimensional parameters (a,b). In non-relativistic point mechanics, a typical molecule has mass of $10^{-26}$ kg and random velocity of $10^3$ m/s. The uncertainty in the measurement of such velocities can be safely put at order of $10^0$ to $10^{+1}$. Thus the uncertainty in momentum, which is equal to $\sqrt{b\epsilon}$ is of the order of $10^{-26}$ to $10^{-25}$. Using the uncertainty principle result of $ab\epsilon^2=\hbar^2$, we find $a\epsilon$ is of the order of $10^{-18}$ to $10^{-16}$ and the uncertainty in the measurement of the coordinate is equal to $10^{-9}$ to $10^{-8}$ m, the size of the molecule. At the atomic quantum level, the mean lifetime of a state is of the order of nanoseconds and must be longer for molecular transitions, thus a time slicing in point mechanical processes at the order of $10^{-6}$ to $10^{-7}$ should already be small giving a of the order of $10^{-11}$. The parameter b must then be of the order of $10^{-40}$, its really small value primarily due to the smallness of the molecular mass. Both parameters are way too small compared to unity although having a broad range of infinitesimal values. Furthermore, they yield uncertainty measurements that are physically meaningful and taking the limit to zero for these infinitesimal values yield the relevant classical equations of motion. Thus, equation (61) reduces to 
\begin{equation}\label{sha62}
\textbf{G}_0=\left(\dfrac{\omega}{2\pi \sin(\omega t)}\right)\delta(\textit{F}_1) \delta(\textit{F}_2).
\end{equation}	   
Using the functional determinant rule
\begin{equation}\label{sha63}
\delta(\textit{F}_1)\delta(\textit{F}_2)=\dfrac{\delta(\chi_1)\delta(\chi_2)}{det\Theta},
\end{equation}
where the functional determinant $\Theta$ is the determinant of the matrix given below and should be evaluated at $\chi_1=\chi_2=0$.
\begin{equation}\label{sha64}
\Theta=\begin{pmatrix}
\dfrac{\delta \textit{F}_1}{\delta \chi_1}  &  \dfrac{\delta \textit{F}_1}{\delta \chi_2} \\
\dfrac{\delta \textit{F}_2}{\delta \chi_1}  &  \dfrac{\delta \textit{F}_2}{\delta \chi_2} \\
\end{pmatrix}.
\end{equation}
Thus, using (53) and  (63) in (62), we see the Wigner kernel for the harmonic oscillator given by Equation (22b). We just need to argue that the extra terms given in equation (43), (59a) and (63) can be absorbed in normalization factor. This is the term
\begin{equation}\label{sha65}
\textbf{N}=(\frac{1}{\epsilon^2})^{\frac{N+1}{2}}(det\left(\textbf{1}+M_0^{-1}M_c\right))^{-\frac{1}{2}}(det\Theta)^{-1},
\end{equation} 
where $\epsilon \rightarrow 0$, $N \rightarrow \infty$, each term in the matrix $\Theta \rightarrow 0$ and $det\left(\textbf{1}+M_0^{-1}M_c\right)$ is order 1.  

	Doing the same thing for the free particle, there are similar steps mirroring equations (44) to (65) with the exception of the absence of equations similar to equations (59a) and (59b) because the matrix M corresponding to equation (57) can be exactly diagonalized with diagonal terms $-\frac{1}{a}\dfrac{d^2}{dt^2}$ and $-\frac{1}{b}\dfrac{d^2}{dt^2}$. Corresponding to equations (60), (53a) and (53b) are
\begin{subequations}\label{sha66} 
\begin{gather}
det^{-\frac{1}{2}}\left(-\frac{1}{a}\dfrac{d^2}{dt^2}\right)=a^{\frac{N}{2}}\left(\frac{1}{2\pi t}\right)^{\frac{1}{2}},\label{first}\\
\chi_1=p+\frac{m}{t}(x-x') \approx 0,\label{second}\\
\chi_2=p'+\frac{m}{t}(x-x') \approx 0.
\end{gather}
\end{subequations}
Using these equations in the corresponding equations to equations (62) to (64), we find the Wigner kernel for the free paerticle given by equation (22a).

	This completes the alternative derivation of the Wigner kernel for the free particle and the harmonic oscillator.

\section{\label{sec:level6}Conclusion}
	In this paper, we presented the path-integral solutions to the Liouville and Wigner equations. The solution requires solving for the Wigner and Liouville kernels and we presented two expressions for these given by equations (14) and (18b). The main part of the paper is to show that these two expressions are identical perturbatively. The perturbative solution needs the Wigner kernel for exact problems like the free particle and harmonic oscillator. These are known in the literature and the answers we gave in equations (22a) and (22b) are derived using equation (18b). We also presented an alternative way of deriving these exact kernels in Section V.
	
	Since the paper addresses how to solve for the distribution, the natural next step is to apply the method. One possibility is to apply it to transport problems, say in quantum electronics. Another is to apply the method to solving the distribution function of a realistic potential like the van der Waal problem. An immediate interesting question is to understand the link between the solutions for the exact kernels given by equations (22a) and (22b). These kernels were solved in earlier papers using canonical transformations, then using the quantum mechanics kernel using equation (18b) and finally directly by path-integral as shown in Section V.
\begin{acknowledgments}
The work of Cilicia Perez was done while she was an MS student at the University of the Philippines, Diliman.
\end{acknowledgments}	 

\end{document}